\documentclass[9pt]{IEEEtran}
\usepackage{cite}
\usepackage{amsmath,amssymb,amsfonts}
\usepackage{algorithmic}
\usepackage{graphicx}
\usepackage{textcomp}
\usepackage{makecell}
\usepackage{booktabs}
\usepackage{xcolor}

\begin{document}

\title{Adaptive Linearly Constrained Minimum Variance Framework for Volumetric Active Noise Control}

\author{
\IEEEauthorblockN{Manan Mittal\IEEEauthorrefmark{2}, Ryan M. Corey\IEEEauthorrefmark{3}, Andrew C. Singer\IEEEauthorrefmark{2}}

\IEEEauthorblockA{\IEEEauthorrefmark{2}
Stony Brook University}
\IEEEauthorblockA{\IEEEauthorrefmark{3}
University of Illinois Chicago and Discovery Partners Institute}
}

\newtheorem{theorem}{Theorem}[section]
\newtheorem{corollary}{Corollary}[theorem]
\newtheorem{lemma}[theorem]{Lemma}
\maketitle

\begin{abstract}  
Traditional volumetric noise control typically relies on multipoint error minimization to suppress sound energy across a region, but offers limited flexibility in shaping spatial responses. This paper introduces a time-domain formulation for linearly constrained minimum variance active noise control (LCMV-ANC) for spatial control filter design. We demonstrate how the LCMV-ANC optimization framework allows system designers to prioritize noise reduction at specific spatial locations through strategically defined linear constraints, providing a more flexible alternative to uniformly weighted multipoint error minimization. An adaptive algorithm based on filtered-X least mean squares (FxLMS) is derived for online adaptation of filter coefficients. Simulation and experimental results validate the proposed method’s noise reduction and constraint adherence, demonstrating effective, spatially selective and broadband noise control compared to multipoint volumetric noise control.
\end{abstract}

\begin{IEEEkeywords}
Active Noise Control, Beamforming, Optimization, Adaptive Filters, Sensor Array Processing
\end{IEEEkeywords}

\section{Introduction}
\label{sec:introduction}

Acoustic noise presents a significant challenge across a wide range of environments, including industrial settings, transportation systems, consumer electronics, and communication devices. The dynamic nature of real-world acoustic environments necessitates the use of adaptive algorithms in active noise control (ANC) systems. Noise sources may vary over time, and the acoustic paths between sources, sensors, and actuators can change dynamically. Adaptive filters address these challenges by continuously updating their parameters to track such variations, thereby maintaining effective noise cancellation. The Filtered-x Least Mean Square (FxLMS) algorithm is widely used in adaptive feedforward ANC systems \cite{sigcont, survey1}.

Volumetric noise control aims to reduce unwanted sound energy throughout a three-dimensional space \cite{3danc, 10887700}, rather than at a limited number of microphone locations. Unlike conventional ANC approaches that minimize noise at specific spatial points (e.g., near the ears or at sensor positions), volumetric control seeks to attenuate noise over an extended region—such as an entire cabin, room, or workspace. This goal typically requires a spatially distributed arrangement of microphones and speakers. Prior methods have attempted to achieve this through multi-point FxLMS control, kernel-interpolation techniques to estimate the acoustic field within a volume \cite{brunnstrom2021kernel}, or virtual sensing strategies \cite{virt_sensing_anc1, benois2022optimization}.

\begin{figure}[htbp]
    \centering
    \includegraphics[width=\linewidth]{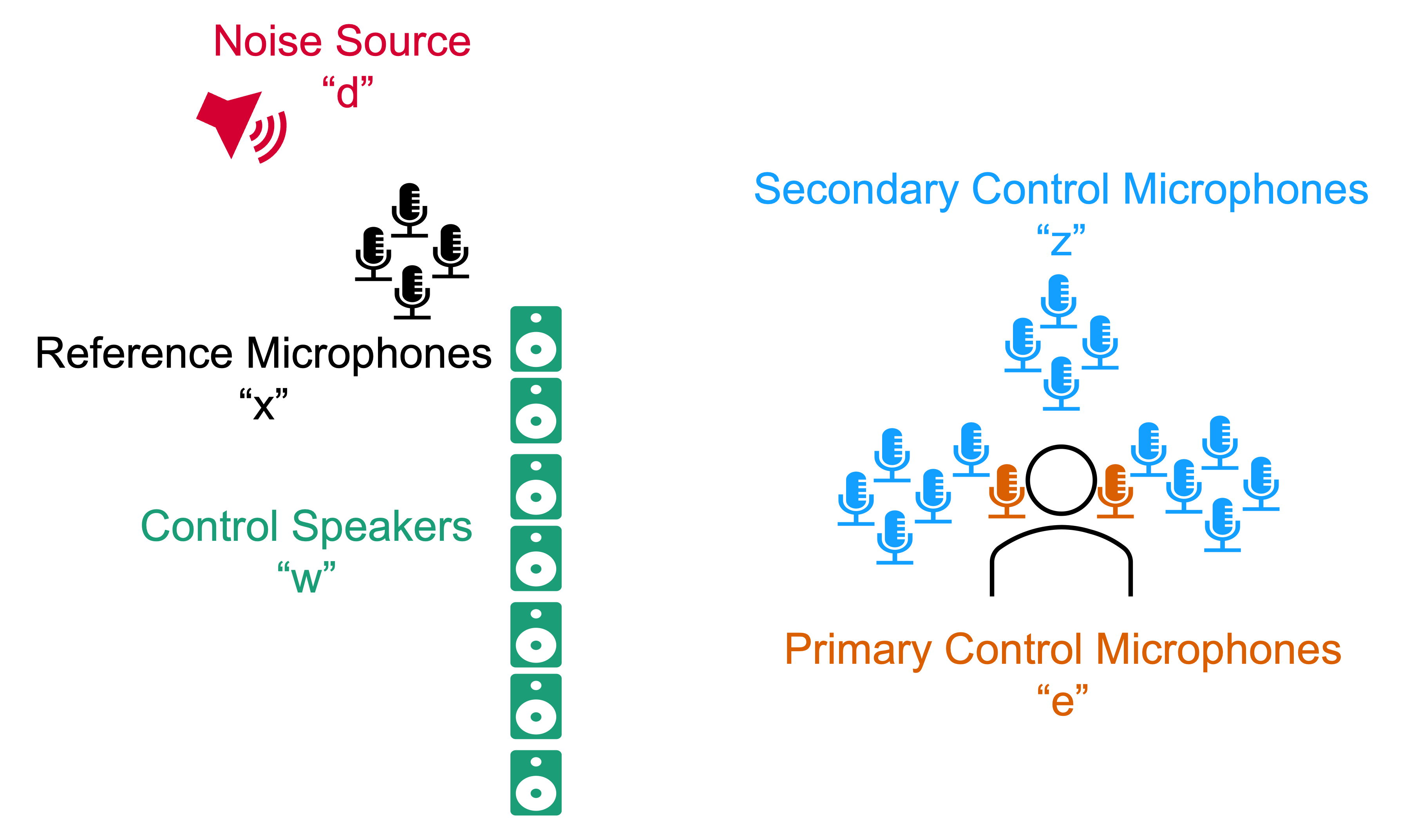}
    \caption{A pictorial depiction of a scenario in which the proposed method may be used. The user has a set of primary and secondary control microphones and the active noise control system aims to reduce the energy in the sound field in that volume while enforcing the field be as small as possible at the user's ears.}
    \label{fig:scenario}
\end{figure}

In array signal processing \cite{johnson1992array}, constrained optimization techniques \cite{boyd2004convex} have proven highly effective, especially for beamforming tasks. The Linearly Constrained Minimum Variance (LCMV) beamformer, in particular, minimizes the output power (variance) of an array while satisfying linear constraints on its response \cite{johnson1992array, lcmvbenny}. These constraints are typically designed to preserve signals from desired directions or suppress interference from known sources. Although LCMV methods are primarily used in spatial filtering with sensor arrays, the core idea—minimizing variance subject to linear constraints—can be adapted to design temporal adaptive filters for ANC. This paper proposes a novel time-domain adaptive algorithm for volumetric ANC, framed within the beamforming paradigm \cite{olivieri2016theoretical}. A frequency-domain LCMV formulation for ANC was introduced by Huang et al.~\cite{lcmv_anc_og}, who derived an ideal solution under the assumption of known primary and secondary paths using acausal, infinite-length filters. However, the assumption of a known primary path is often impractical. The method presented in this work eliminates this assumption and learns the necessary information directly from the data.

A key advantage of the proposed algorithm is its operation entirely in the time domain \cite{brunnstrom2025time}, which is beneficial in latency-sensitive applications where frequency-domain methods may introduce delay. By combining the LCMV optimization objective with the low-latency, time-domain generalized sidelobe canceler structure, the proposed algorithm provides effective volumetric noise control with explicitly enforced linear constraints. Furthermore, the method naturally extends to kernel-interpolated fields. This provides a flexible design trade-off: Certain spatial locations—such as those near a listener’s ears—can be strictly constrained to achieve reliable noise attenuation, while the rest of the volume can be optimized under a minimum variance objective. In contrast to prior works—such as the dual zone formulation \cite{app10010004}, the directional ANC \cite{10289757, 8917683} with minimax \cite{app9194065}, and spatially selective systems to preserve certain signals \cite{su2024spatial, xiao2023spatially, xiao2025soft, xiao2025spatially, xiao2024effect}, a fundamentally different problem, we pose the problem for controlled filter design in a volume which may be coupled with directional or spatially selective constraints and extends to an arbitrary number of zones.

This hybrid approach ensures strong performance at critical points without sacrificing global noise reduction. Simulation and experimental results demonstrate the effectiveness of the proposed adaptive LCMV-ANC approach.

\section{Signal Model}

\begin{table}[htbp]
\centering
\caption{List of symbols}
\begin{tabular}{ll}
\hline
\textbf{Symbol} & \textbf{Description} \\
\hline
$t$ & Time \\
$N_s$ & Number of control speakers \\
$N_e$ & Number of primary control microphones\\
$N_z$ & Number of secondary control microphones\\
$N_r$ & Number of reference microphones\\
$N_t$ & Number of filter co-efficients \\
$p_e[t]$ & Primary path impulse response to primary control \\
& microphones\\
$p_z[t]$ & Primary path impulse response to secondary control \\
&microphones\\
$g_{e,s}[t]$ & Secondary path impulse response to primary control \\ & microphones\\
$\mathbf{X}_e[t]$ & Reference microphone signals filtered by impulse \\ &responses for primary control microphones\\
$g_{z,s}[t]$ & Secondary path impulse response to secondary control \\ &microphones\\
$\mathbf{X}_z[t]$ & Reference microphone signals filtered by impulse \\ &responses to secondary control microphones\\
$\mathbf{w}$ & Filter weights\\
\hline
\end{tabular}
\label{los}
\end{table}
\noindent
Suppose we have an acoustic environment with $N_s$ control speakers and $N_e$ primary control microphones. To monitor the acoustic environment, there are $N_r$ reference microphones. All sensors are distributed in the environment without an assumed geometry. The noise source generates a sequence $d$ over time. Table \ref{los} includes the list of the remaining symbols used in the signal model. The received signals at the error microphones $e_j, j \in [1 , \ldots, N_e]$ at time $t$ are given by,
\begin{equation}
e_j[t] = (p_{e_j} \ast d)[t] + \Sigma_{s=1}^{N_s} \Sigma_{r=1}^{N_r} (g_{e_j,s} \ast w_{s,r} \ast x_r)[t].
\end{equation}
Here, $\ast$ denotes convolution. Similarly, the received signals at the $N_z$ secondary control locations $z_k, k \in [1, \ldots, N_z]$ are given by, 
\begin{equation}
z_k[t] = (p_{z_k} \ast d)[t] + \Sigma_{s=1}^{N_s} \Sigma_{r=1}^{N_r} (g_{z_k,s} \ast w_{s,r} \ast x_r)[t].
\end{equation}
The received signals may be represented in matrix-vector notation. Let,
\begin{align*}
\mathbf{e}[t]&=[e_1[t], e_2[t], \ldots, e_{N_e}[t]]^T \in \Re^{N_e} \\
\mathbf{d_e}[t] &= [(p_{e_1} \ast d)[t],\ldots, (p_{e_{N_e}} \ast d)[t]]^T \in \Re^{N_e} \\
\mathbf{w} &= \Big[
w_{1,1}[0] \;\; w_{1,2}[0] \;\; \cdots \;\; w_{1,N_r}[0] \;\; \cdots \\
&\qquad w_{N_s,N_r}[0] \;\; \cdots \;\; w_{N_s,N_r}[N_t - 1]
\Big]^T \in \Re^{N_sN_rN_t}, \\
\mathbf{X_e}[t] &=
\begin{bmatrix}
\tilde{\mathbf{x}}_1^T[t] & \tilde{\mathbf{x}}_1^T[t-1] & \cdots & \tilde{\mathbf{x}}_1^T[t - N_t + 1] \\
\vdots & \vdots & \ddots & \vdots \\
\tilde{\mathbf{x}}_{N_e}^T[t] & \tilde{\mathbf{x}}_{N_e}^T[t-1] & \cdots & \tilde{\mathbf{x}}_{N_e}^T[t - N_t + 1]
\end{bmatrix}, \\
\tilde{\mathbf{x}}_e[t] &= \Big[
\tilde{x}_{e,1,1}[t] \;\; \tilde{x}_{e,1,2}[t] \;\; \cdots \;\; \tilde{x}_{e,1,N_r}[t] \;\; \tilde{x}_{e,2,1}[t] \;\; \cdots \\
&\qquad \tilde{x}_{e,N_s,N_r}[t]
\Big]^T, \\
\tilde{x}_{e,s,r}[t] &= (g_{e,s} \ast x_r)[t].
\end{align*}
Therefore, 
\begin{align}
\mathbf{e}[t] &= \mathbf{d_e}[t] + \mathbf{X_e}[t]\mathbf{w} \\
\mathbf{z}[t] &= \mathbf{d_z}[t] + \mathbf{X_z}[t]\mathbf{w},
\end{align}
where $\mathbf{z}[t], \mathbf{d_z}[t]$ and $\mathbf{X_z}[t]$ are constructed in the same manner as $\mathbf{e}[t], \mathbf{d_e}[t]$ and $\mathbf{X_e}[t]$ but instead making use of impulse responses to the secondary control locations. Note, this form of the received signals is useful in the derivation of various filtered-X algorithms used in active noise control. In the next section, we review the FxLMS algorithm. We present the multi-point and two-point optimization problems that we compare against in this work.

\section{Filtered-x Least Mean Squares (FxLMS) Active Noise Control}
Filtered-x Least Mean Squares (FxLMS) is a widely-used adaptive algorithm in active noise control, designed to account for the secondary path dynamics between the control source and the error sensors. Unlike traditional LMS, FxLMS filters the reference signal through an estimate of the secondary path before adaptation. This compensates for phase and magnitude distortions, ensuring stable convergence. In this paper, we compare the performance of the proposed method to multi-point FxLMS to jointly minimize the error at both primary and secondary control locations and two-point FxLMS to minimize the error only at the primary control locations. We first present the optimization used for two-point FxLMS and the multi-point variant after.

\subsection{Two-point FxLMS}
In the case of two-point FxLMS, the optimization problem can be presented as follows,
\begin{align*}
    J(\mathbf{w}) = \min_{\mathbf{w}} \; E[\mathbf{e}[t]^T \mathbf{e}[t]]
\end{align*}
where $E$ denotes expectation. In a real-time scenario, the update rule for two-point FxLMS is given by
\begin{align}
    \mathbf{w}[t] = \mathbf{w}[t-1] - \alpha \mathbf{X}_e^T[t] \mathbf{e}[t]
\end{align}
Here, $\alpha$ is the learning rate. 

\subsection{Multi-point FxLMS}
In the case of multi-point FxLMS, the optimization problem can be presented as follows.
\begin{align*}
    J(\mathbf{w}) = \min_{\mathbf{w}} \; E[\begin{bmatrix}
\mathbf{e}[t]^T & \mathbf{z}[t]^T
\end{bmatrix}
\begin{bmatrix}
\mathbf{e}[t] \\
\mathbf{z}[t]
\end{bmatrix}]
\end{align*}
In a real-time scenario, the update rule for multi-point FxLMS is given by
\begin{align}
    \mathbf{w}[t] = \mathbf{w}[t-1] - \alpha 
        [\mathbf{X}_e^T[t] \mathbf{X}_z^T[t]]
    \begin{bmatrix}
        \mathbf{e}[t] \\
        \mathbf{z}[t]
    \end{bmatrix}
\end{align}
Here, $\alpha$ is the learning rate. FxLMS is widely adopted due to its simplicity and robustness, and serves as a baseline for the proposed method. One may used normalized FxLMS where the step is normalized by the input energy at the control locations. The next section introduces the proposed time domain formulation for LCMV-ANC and the corresponding adaptive algorithm. 

\section{Linearly Constrained Minimum Variance Active Noise Control}
Linearly constrained minimum variance active noise control aims to perform informed spatial filter design for volumetric ANC. Traditional volumetric multi-point FxLMS or kernel-FxLMS algorithms aim to minimize the error power at all locations with uniform weighting for each constraint location. When position estimates can be made, as in the case of head-tracked ANC \cite{nature_elliot}, LCMV-ANC allows the system designer to trade-off constraint adherence at known locations and regional noise control in the desired volume. 

\subsection{Optimization Problem}
We present an altered form of the optimization first proposed by Huang, et al. \cite{lcmv_anc_og}. At each time t, we seek filter weights such that they minimize the cost function given by,
\begin{align*}
J(\mathbf{w}) = \min_{\mathbf{w}}& \; \frac{\mathbf{z}[t]^T \mathbf{z}[t]}{2} \\
\text{subject to} \quad &\mathbf{e}[t] = \mathbf{0}.
\end{align*}
The time parameter $t$ is omitted for brevity henceforth. Let $\mathbf{A}=(\mathbf{X_z}^T\mathbf{X_z} + \epsilon\mathbf{I}_{{N_s}{N_r}{N_t}})^{-1}$, where $\epsilon$ is a regularization parameter to ensure stability. The optimal filter weights $\mathbf{w}_\mathrm{opt} = \arg \min_\mathbf{w} J(\mathbf{w})$ are given by \cite{lcmv_anc_og},
\begin{align*}
    \mathbf{w}_\mathrm{opt} = &( \mathbf{A}
    \mathbf{X_e}^T(\mathbf{X_e}\mathbf{A}\mathbf{X_e}^T + \mu\mathbf{I}_{N_e})^{-1} \\
    &(\mathbf{X_e}\mathbf{A}\mathbf{X_z}^T\mathbf{d_z} - \mathbf{d_e})) - \mathbf{A}\mathbf{X_z}^T\mathbf{d_z}.
\end{align*}
Here, $\mu$ is another regularization parameter to ensure stability in the inversion. Here, the regularization parameters are added as stable approximations of the true solution.

\subsection{Adaptive Algorithm}
In a real-time active noise control scenario, one must update the filter coefficients online to adapt to the time-varying noise. We apply stochastic gradient descent (SGD) to the Lagrangian in the optimization problem. Explicitly, the update rule may be expressed as,
\begin{align}
    \mathbf{w}[t] &= \mathbf{w}[t-1] - \alpha \nabla_w(\frac{\mathbf{z}^T\mathbf{ z}}{2} + \mathbf{\lambda}^T \mathbf{e}) \\
    &= \mathbf{w}[t-1] - \alpha(\mathbf{X_z}^T\mathbf{z} + \mathbf{X_e}^T\mathbf{\lambda}).
\end{align}
Here, $\alpha$ is the learning rate. To solve for $\mathbf{\lambda}$ we must enforce the linear constraints. Substitution of the update equation, into the linear constraint $\mathbf{e} = 0$, results in the following expression,
\begin{equation} \label{lambda}
    \mathbf{\lambda} = -(\mathbf{X_e}\mathbf{X_e}^T)^{-1}(\mathbf{X_e}\mathbf{X_z}^T\mathbf{z} - \frac{\mathbf{e}}{\alpha})
\end{equation}
This yields the final expression for the weight update,
\begin{equation}
\begin{split}
    \mathbf{w}[t] = &\mathbf{w}[t-1] - \alpha((\mathbf{I}_{{N_s}{N_r}{N_t}} -\mathbf{X_e}^T(\mathbf{X_e}\mathbf{X_e}^T)^{-1}\mathbf{X_e})\mathbf{X_z}^T\mathbf{z}) - \\
    & \mathbf{X_e}^T(\mathbf{X_e}\mathbf{X_e}^T)^{-1}\mathbf{e} \\
    = &\mathbf{w}[t-1] - \alpha \mathbf{P^C_{X_e}\mathbf{X_z}^T\mathbf{z}} - \mathbf{P_{X_e}^{MN}\mathbf{e}}
\end{split}
\end{equation}
where $\mathbf{P_{X_e}^C}$ is the linear projector into the kernel (nullspace) of $\mathbf{X_e}[t]$ and $\mathbf{P^{MN}_{X_e}}$ is the linear minimum norm projector of $\mathbf{X_e}$. 

\section{Results and Discussion}
\subsection{Simulation Results}\label{sim}
\begin{figure}[htbp]
    \centering
    \includegraphics[width=\linewidth, height=5cm]{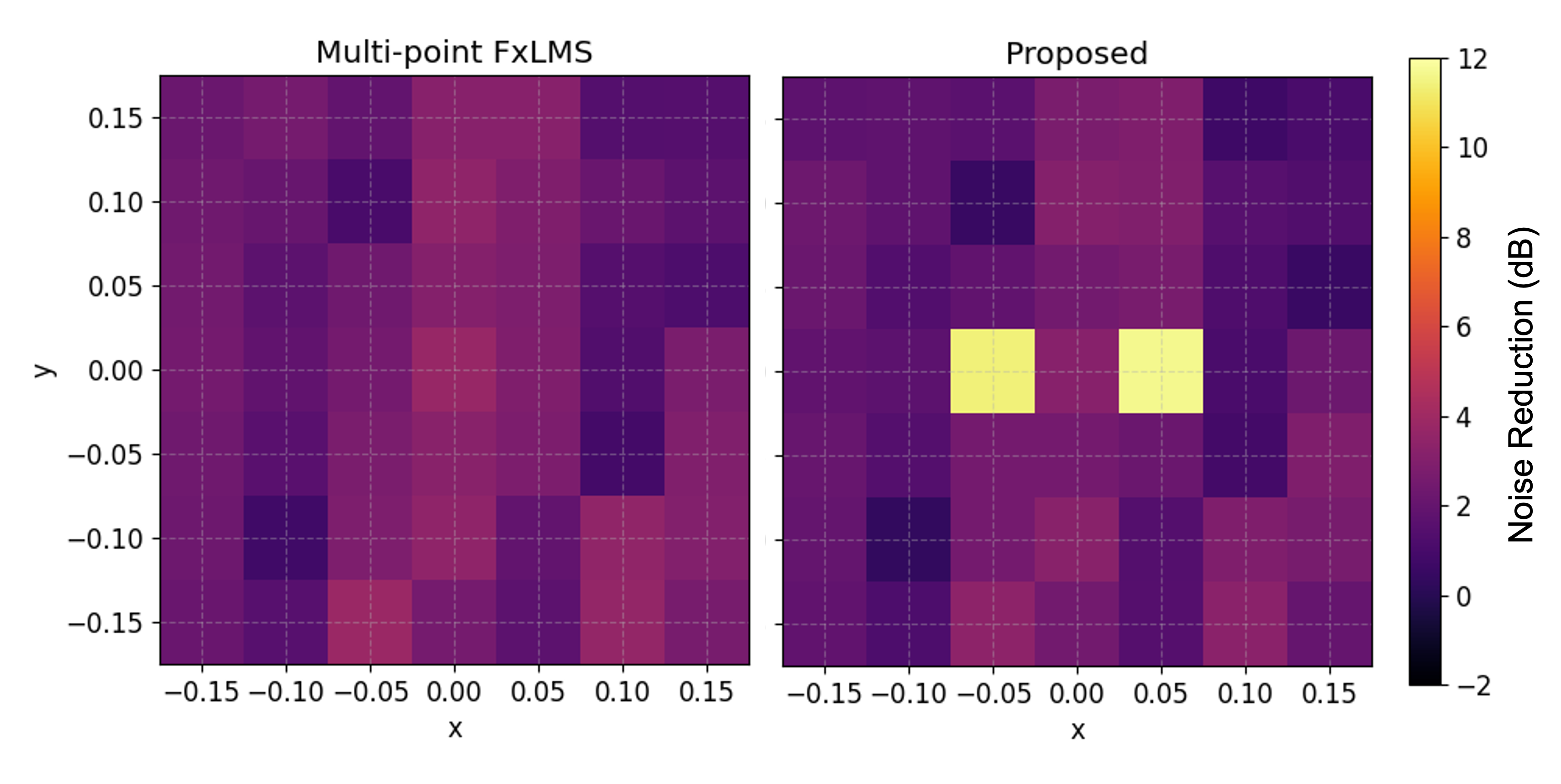}
    \caption{Noise reduction in dB for the proposed method and multipoint FxLMS variants over the volume in section \ref{sim}. The proposed method is able to outperform multi-point FxLMS at both the primary control locations while being marginally worse over the entire volume.}
    \label{fig:meshrir_heatmap}
    \vspace{-2mm}
\end{figure}

The method is evaluated in a simulated environment using impulse responses from the MeshRIR S32-M441 dataset \cite{MeshRIR}. A loudspeaker positioned at (-0.7$\mathrm{m}$, -1$\mathrm{m}$, 0.1 $\mathrm{m}$) is used as the noise source, while the remaining 31 loudspeakers serve as control sources. The negative coordinates arise due to the reference origin being set at the center of the room. In this simulation, we employ 2 primary control locations and 60 secondary control locations, which together span slightly less than 0.16 $\mathrm{m^2}$ as defined in the MeshRIR dataset.
We assume access to the true noise signal, which is generated by sampling a zero-mean, unit-variance Gaussian distribution. Having access to a reference microphone is equivalent to having the noise signal convolved with a (possibly unknown) reference path. Mathematically, it introduces no fundamental change to the model. Our experiments therefore remain general and physically consistent. The performance of the proposed adaptive filtering method is compared against the standard multi-point FxLMS algorithm. Additionally, we include a variant of the multi-point FxLMS constrained to minimize power at the primary control locations only.
The simulated signal is played for 120 seconds, with filter coefficients updated at every sample. Due to the multiple constraints imposed, the optimal solution is underdetermined. 
Figure~\ref{fig:meshrir_heatmap} displays a heatmap of noise reduction across the central region of the volume of interest, further highlighting the advantage of the proposed method. It achieves noise reduction within 0.5$\,$dB of the multi-point FxLMS at all secondary control locations, while providing superior performance at the primary control points due to better constraint adherence. Notably, the proposed method significantly outperforms the multi-point FxLMS at the constrained locations.

\subsection{Experimental Results} \label{exp}
\begin{figure}[htbp]
    \centering
    \includegraphics[width=\linewidth]{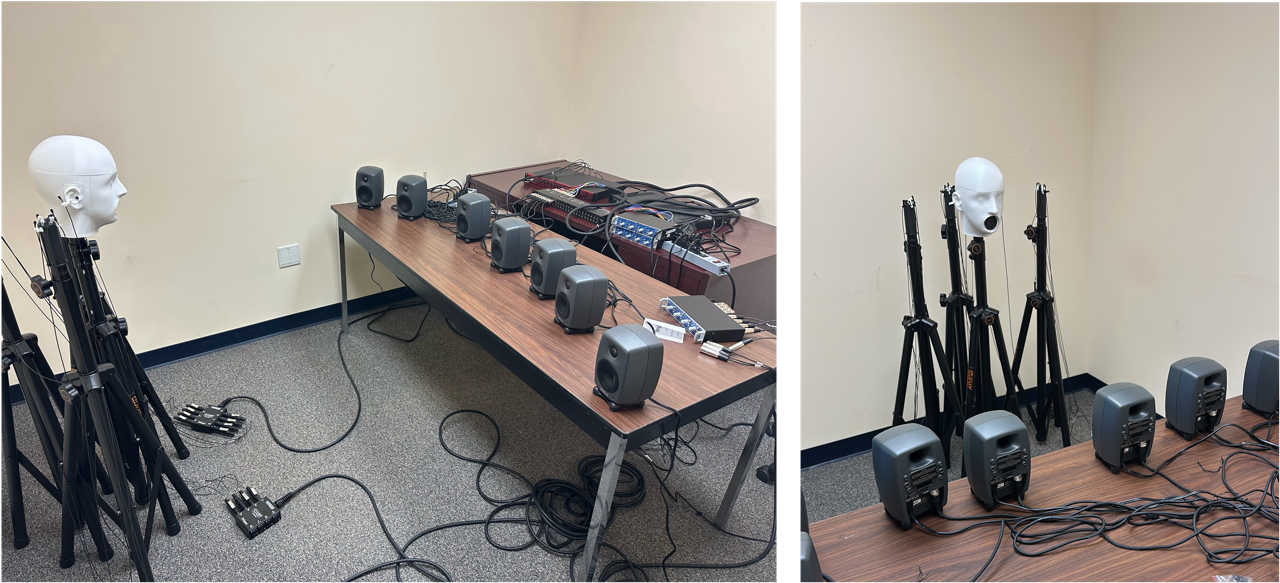}
    \caption{The experimental setup used to test the proposed method in a real-world environment. The acoustic head simulator has two in-ear microphones and 2 attached to the ears. Distributed around the volume are 16 microphones. 7 loudspeakers are used as control speakers. Not shown is the loudspeaker used as a noise source.}
    \label{fig:exp_setup}
\end{figure}

\begin{figure}[htbp]
    \centering
    \includegraphics[width=\linewidth]{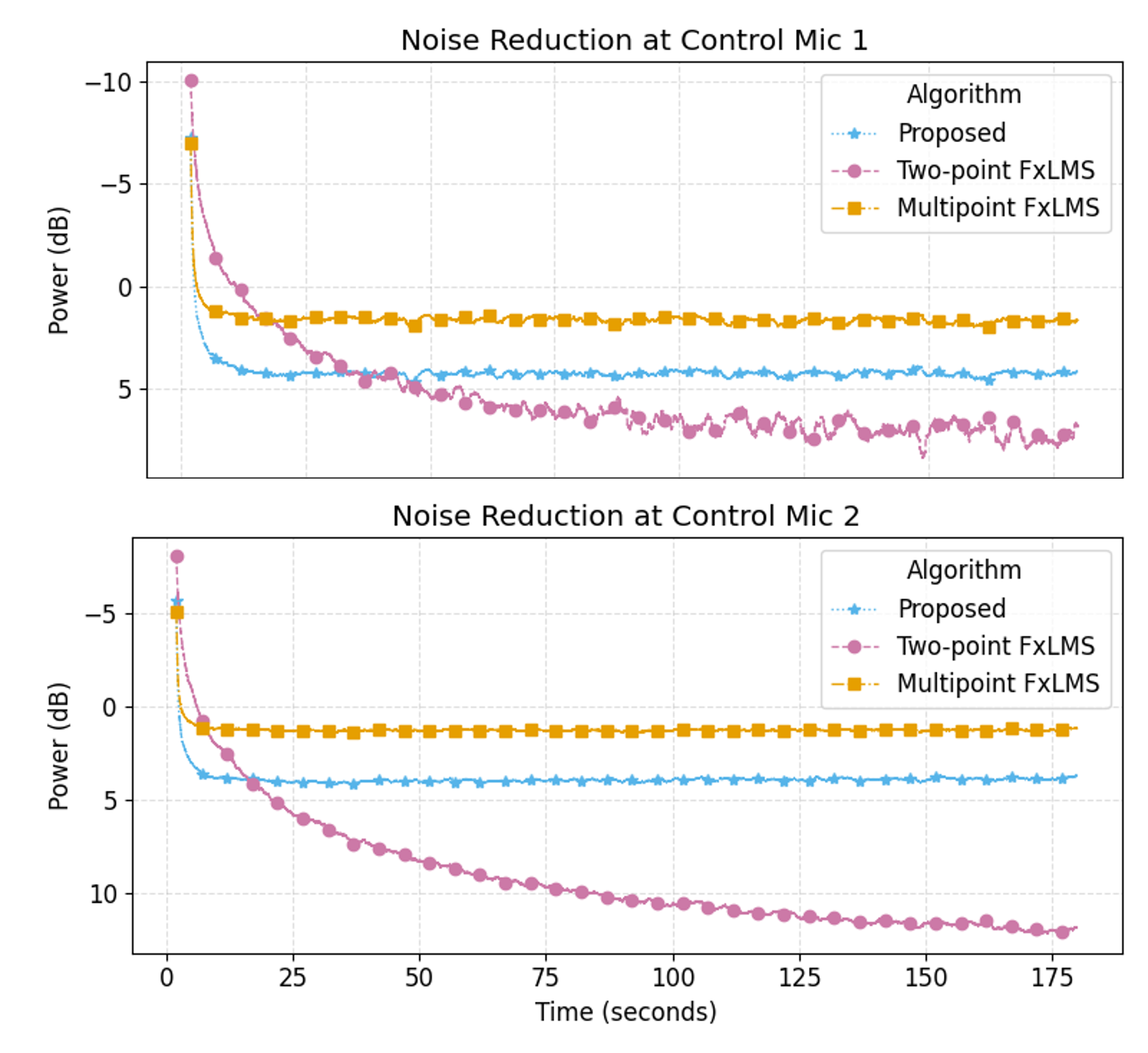}
    \caption{Convergence curves for the proposed method and the two FxLMS variants in the experimental setup of section \ref{exp}. The proposed method is able to outperform multi-point FxLMS at both the primary control locations.}
    \label{fig:sbu_primary_control_locations}
    \vspace{-5mm}
\end{figure}

To evaluate the real-world efficacy of the proposed method, data was collected in a room with a reverberation time of $RT_{60} \approx 490\,\mathrm{ms}$. Figure~\ref{fig:exp_setup} illustrates the experimental setup, while Figure~\ref{fig:scenario} provides a depiction of the same. The setup includes seven control loudspeakers and one primary noise source (not pictured). The acoustic head simulator \cite{heads}, shown in the figure, is equipped with two in-ear microphones, which were used as the primary control locations in this experiment, along with two additional microphones positioned at the ears. In addition, 16 microphones are distributed around the acoustic head simulator to capture acoustic information over a larger spatial volume surrounding the control locations.
A zero-mean, unit-variance white noise signal is played back through the primary noise source. All 20 microphone signals are recorded synchronously at a sampling rate of 48 kHz. Impulse responses from each of the control loudspeakers are measured and used in the filtered-X algorithms. As in the simulated case, it is assumed that the system designer has access to the original noise sequence.

\begin{figure}[htbp]
    \centering
    \includegraphics[width=\linewidth]{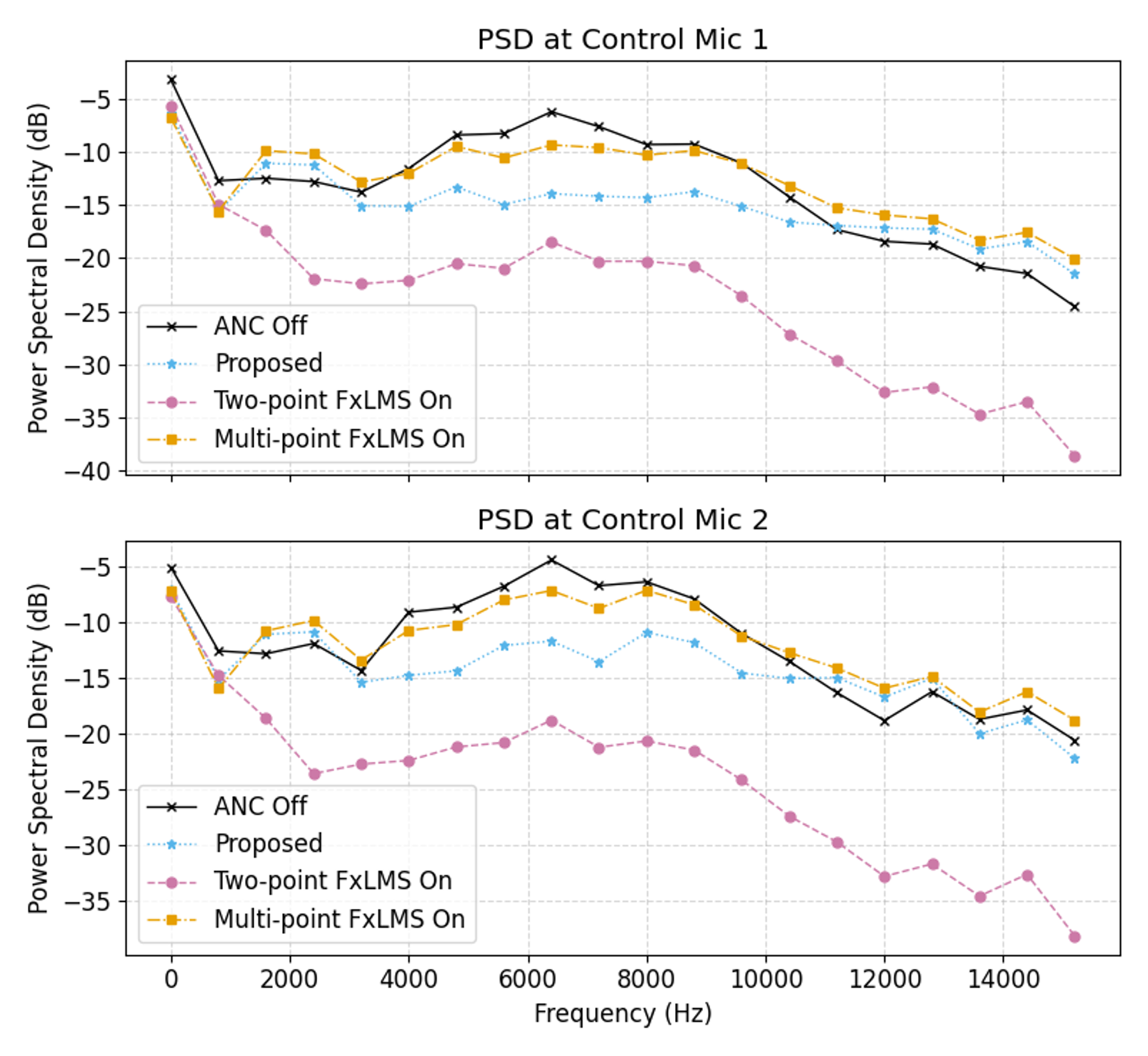}
    \caption{Power spectral density in dB for the proposed method and the two FxLMS variants in the experimental setup of section \ref{exp}. The proposed method is able to outperform multi-point FxLMS at both the primary control locations.}
    \label{fig:sbu_primary_control_locations_psd}
\end{figure}

\begin{figure}[htbp]
    \centering
    \includegraphics[width=\linewidth]{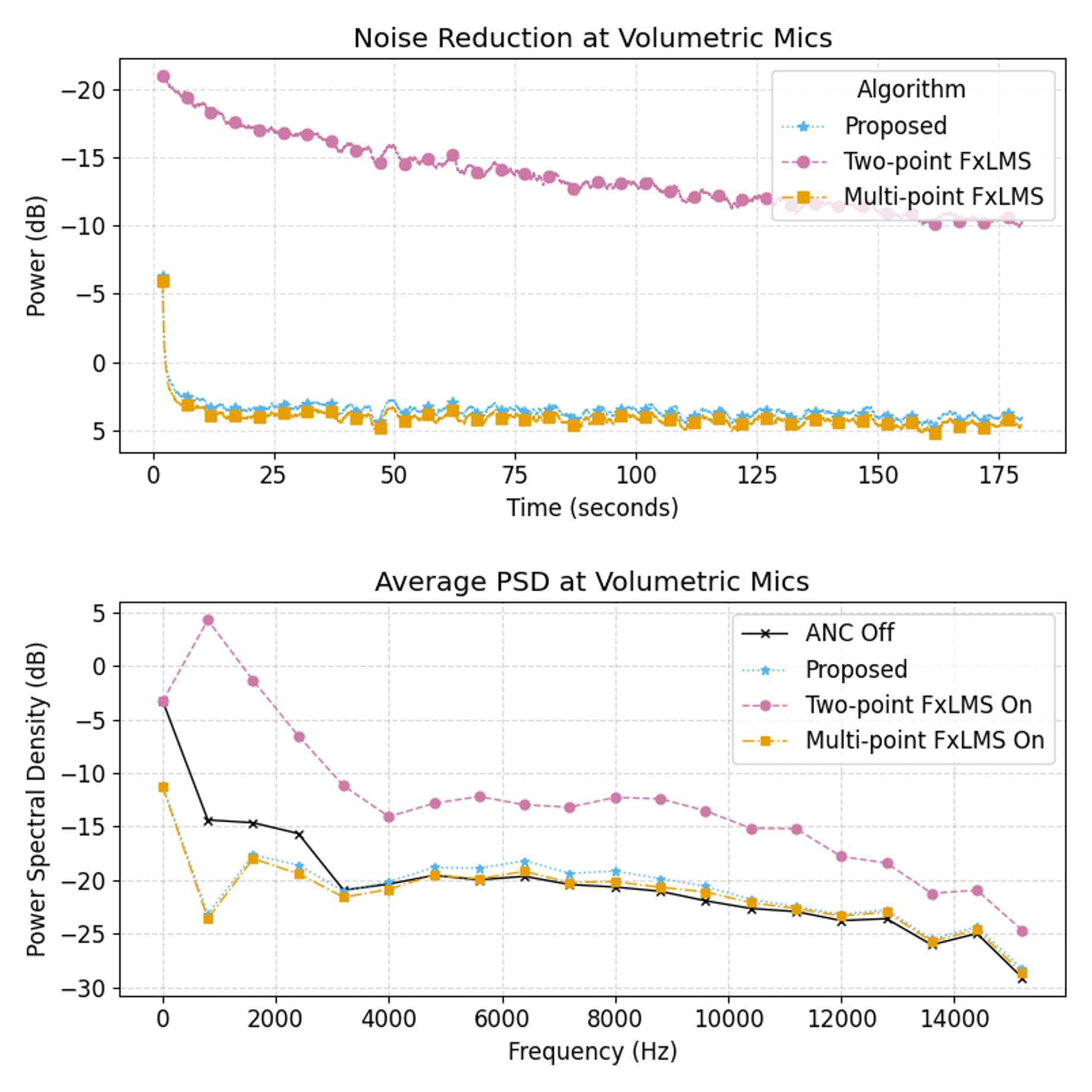}
    \caption{Average convergence curves for the proposed method and the two FxLMS variants in the experimental setup of section \ref{exp} for the secondary control mics. The power spectral density in dB shows that the proposed method is able to perform comparably to multi-point FxLMS.}
    \label{fig:sbu_avg}
    \vspace{-5mm}
\end{figure}
Figure \ref{fig:sbu_primary_control_locations} shows the convergence curves for the proposed method, two-point FxLMS and multi-point FxLMs. The proposed method is able to reduce noise at the constrained location more effectively than multi-point FxLMS but is unable to match two-point FxLMS. Figure \ref{fig:sbu_primary_control_locations_psd} shows the power spectral density (PSD), plotted in decibels (dB), at the two in-ear microphones for each algorithm. The PSD is limited to the 0–16 kHz frequency range for clarity. The proposed method clearly achieves better satisfaction of the linear constraints compared to the multi-point FxLMS, with noticeably improved performance across a broad frequency range. The reduced low frequency performance is because the acoustic head simulator's HRTF must be estimated along with the room transfer function for those points causing a significant change in the acoustic manifold in that region. Although the proposed method trades off some performance at secondary control locations due to the inclusion of a minimum variance component, it still achieves results within 1 dB of the multi-point FxLMS at those locations.
Figure \ref{fig:sbu_avg} shows the average convergence behavior and the average PSD (also in dB, limited to 0–16 kHz) at the secondary control microphones. These results demonstrate that the proposed method delivers superior performance at the primary control locations, while maintaining performance at the secondary locations comparable to that of the multi-point FxLMS.

\section{Conclusion}
The time-domain formulation and adaptive algorithm presented in this work reframe active noise control as a problem of informed spatial trade-offs rather than uniform suppression. By embedding linear constraints directly into the optimization process via a Lagrangian approach, our method allows precise control over where in space noise reduction is prioritized, reflecting a shift from purely reactive to strategically targeted spatial filter design. The demonstrated gains at constrained locations without degrading overall performance affirm the value of this perspective. This framework of LCMV active noise control allows system designers flexibility in how degrees of freedom are allocated for the control of the acoustic field. In particular, our method readily extends to any spatial filtering scheme for active noise control and allows for effective spatial control, particularly when coupled with future extensions such as kernel-based noise field interpolation, which promises to further reduce the burden of spatial sampling while enhancing control fidelity.

\bibliography{refs}

\begin{thebibliography}{10}

\bibitem{sigcont}
S.~Elliott, {\em Signal Processing for Active Control}.
\newblock 01 2001.

\bibitem{survey1}
L.~Lu, K.-L. Yin, R.~C. {de Lamare}, Z.~Zheng, Y.~Yu, X.~Yang, and B.~Chen, ``A survey on active noise control in the past decade---part i: Linear systems,'' {\em Signal Processing}, vol.~183, p.~108039, 2021.

\bibitem{3danc}
I.~Ardekani and W.~Abdulla, ``Active noise control in three dimensions,'' {\em Control Systems Technology, IEEE Transactions on}, vol.~22, pp.~2150--2159, 11 2014.

\bibitem{10887700}
H.~Zhang, H.~J. Sun, J.~A. Zhang, P.~Samarasinghe, and Y.~A. Zhang, ``A spherical-harmonic domain selective spatial active noise control system based on sound field reproduction,'' in {\em ICASSP 2025 - 2025 IEEE International Conference on Acoustics, Speech and Signal Processing (ICASSP)}, pp.~1--5, 2025.

\bibitem{brunnstrom2021kernel}
J.~Brunnstr{\"o}m and S.~Koyama, ``Kernel-interpolation-based filtered-x least mean square for spatial active noise control in time domain,'' in {\em ICASSP 2021-2021 IEEE International Conference on Acoustics, Speech and Signal Processing (ICASSP)}, pp.~161--165, IEEE, 2021.

\bibitem{virt_sensing_anc1}
J.~Zhang, S.~J. Elliott, and J.~Cheer, ``Robust performance of virtual sensing methods for active noise control,'' {\em Mechanical Systems and Signal Processing}, vol.~152, p.~107453, 2021.

\bibitem{benois2022optimization}
P.~R. Benois, R.~Roden, M.~Blau, and S.~Doclo, ``Optimization of a fixed virtual sensing feedback anc controller for in-ear headphones with multiple loudspeakers,'' in {\em ICASSP 2022-2022 IEEE International Conference on Acoustics, Speech and Signal Processing (ICASSP)}, pp.~8717--8721, IEEE, 2022.

\bibitem{johnson1992array}
D.~H. Johnson and D.~E. Dudgeon, {\em Array signal processing: concepts and techniques}.
\newblock Simon \& Schuster, Inc., 1992.

\bibitem{boyd2004convex}
S.~P. Boyd and L.~Vandenberghe, {\em Convex optimization}.
\newblock Cambridge university press, 2004.

\bibitem{lcmvbenny}
M.~Souden, J.~Benesty, and S.~Affes, ``A study of the lcmv and mvdr noise reduction filters,'' {\em IEEE Transactions on Signal Processing}, vol.~58, no.~9, pp.~4925--4935, 2010.

\bibitem{olivieri2016theoretical}
F.~Olivieri, F.~M. Fazi, P.~A. Nelson, M.~Shin, S.~Fontana, and L.~Yue, ``Theoretical and experimental comparative analysis of beamforming methods for loudspeaker arrays under given performance constraints,'' {\em Journal of Sound and Vibration}, vol.~373, pp.~302--324, 2016.

\bibitem{lcmv_anc_og}
S.-C. Huang, C.-H. Ma, Y.-C. Hsu, and M.~R. Bai, ``Feedforward active noise global control using a linearly constrained beamforming approach,'' {\em Journal of Sound and Vibration}, vol.~537, p.~117190, 2022.

\bibitem{brunnstrom2025time}
J.~Brunnstr{\"o}m, M.~B. M{\o}ller, J.~{\O}stergaard, S.~Koyama, T.~van Waterschoot, and M.~Moonen, ``Time-domain sound field estimation using kernel ridge regression,'' {\em arXiv preprint arXiv:2509.05720}, 2025.

\bibitem{app10010004}
R.~Wang, X.~Wang, J.~Liu, and J.~Yang, ``Dual-zone active noise control algorithm,'' {\em Applied Sciences}, vol.~10, no.~1, 2020.

\bibitem{10289757}
H.~Zhang, J.~Zhang, F.~Ma, P.~N. Samarasinghe, and H.~Sun, ``A time-domain multi-channel directional active noise control system,'' in {\em 2023 31st European Signal Processing Conference (EUSIPCO)}, pp.~376--380, 2023.

\bibitem{8917683}
V.~Patel, J.~Cheer, and S.~Fontana, ``Design and implementation of an active noise control headphone with directional hear-through capability,'' {\em IEEE Transactions on Consumer Electronics}, vol.~66, no.~1, pp.~32--40, 2020.

\bibitem{app9194065}
K.~Chi, M.~Wu, R.~Han, C.~Gong, and J.~Yang, ``Directional active noise control with a local minimax error criterion,'' {\em Applied Sciences}, vol.~9, no.~19, 2019.

\bibitem{su2024spatial}
X.~Su, D.~Shi, Z.~Zhu, W.-S. Gan, and L.~Ye, ``Spatial-frequency-based selective fixed-filter algorithm for multichannel active noise control,'' {\em IEEE Signal Processing Letters}, 2024.

\bibitem{xiao2023spatially}
T.~Xiao, B.~Xu, and C.~Zhao, ``Spatially selective active noise control systems,'' {\em The Journal of the Acoustical Society of America}, vol.~153, no.~5, pp.~2733--2733, 2023.

\bibitem{xiao2025soft}
T.~Xiao, R.~Roden, M.~Blau, and S.~Doclo, ``Soft-constrained spatially selective active noise control for open-fitting hearables,'' {\em arXiv preprint arXiv:2507.12122}, 2025.

\bibitem{xiao2025spatially}
T.~Xiao and S.~Doclo, ``Spatially selective active noise control for open-fitting hearables with acausal optimization,'' {\em arXiv preprint arXiv:2505.10372}, 2025.

\bibitem{xiao2024effect}
T.~Xiao and S.~Doclo, ``Effect of target signals and delays on spatially selective active noise control for open-fitting hearables,'' in {\em ICASSP 2024-2024 IEEE International Conference on Acoustics, Speech and Signal Processing (ICASSP)}, pp.~1056--1060, IEEE, 2024.

\bibitem{nature_elliot}
S.~J. Elliott, W.~Jung, and J.~Cheer, ``Head tracking extends local active control of broadband sound to higher frequencies,'' {\em Scientific reports}, vol.~8, no.~1, p.~5403, 2018.

\bibitem{MeshRIR}
S.~Koyama, T.~Nishida, K.~Kimura, T.~Abe, N.~Ueno, and J.~Brunnström, ``Meshrir: A dataset of room impulse responses on meshed grid points for evaluating sound field analysis and synthesis methods,'' in {\em 2021 IEEE Workshop on Applications of Signal Processing to Audio and Acoustics (WASPAA)}, pp.~1--5, 2021.

\bibitem{heads}
A.~Lu, K.~Sarkar, Y.~Zhuang, L.~Lin, R.~M. Corey, and A.~C. Singer, ``Accelerating audio research with robotic dummy heads,'' 2025.

\end{thebibliography}
\vspace{12pt}
\color{red}

\end{document}